# A Hybrid Approach to Web Service Recommendation Based on QoS-Aware Rating and Ranking

Mingming Chen[1,2] and Yutao Ma[2,*]
1. State Key Laboratory of Software Engineering, Wuhan University, Wuhan 430072, China
2. School of Computer, Wuhan University, Wuhan 430072, China
{cmm2010, ytma}@whu.edu.cn
*Corresponding author

**ABSTRACT:**

As the number of Web services with the same or similar functions increases steadily on the Internet, nowadays more and more service consumers pay great attention to the non-functional properties of Web services, also known as quality of service (QoS), when finding and selecting appropriate Web services. For most of the QoS-aware Web service recommendation systems, the list of recommended Web services is generally obtained based on a rating-oriented prediction approach, aiming at predicting the potential ratings that an active user may assign to the unrated services as accurately as possible. However, in some application scenarios, high accuracy of rating prediction may not necessarily lead to a satisfactory recommendation result. In this paper, we propose a ranking-oriented hybrid approach by combining the item-based collaborative filtering and latent factor models to address the problem of Web services ranking. In particular, the similarity between two Web services is measured in terms of the correlation coefficient between their rankings instead of between the traditional QoS ratings. Besides, we also improve the measure NDCG (Normalized Discounted Cumulative Gain) for evaluating the accuracy of the top K recommendations returned in ranked order. Comprehensive experiments on the QoS data set composed of real-world Web services are conducted to test our approach, and the experimental results demonstrate that our approach outperforms other competing approaches.

# INTRODUCTION

Web service has been deemed as a promising technique to provide easy-to-access software functions through standard web protocols, and it aims to ensure effective communication between two electronic devices from the same or a different platform over a network. The past fifteen years have witnessed the increasing popularity of SOA (Service-Oriented Architecture) in the era of Web 2.0 (O'reilly, 2007). With the rapid growth of Web services on the Internet, how to choose the most appropriate Web service(s) for service requestors, especially from a large number of Web services available that have similar or even identical functions, becomes an open challenge to the field of Services Computing (Shao, Zhang *et al*. 2007). Hence, this calls for effective approaches to Web service selection and recommendation.

In Services Computing, QoS (Quality of Service) represents the non-functional properties of Web services (including response time, throughput, availability, security and other attributes), which are important factors for service requestors to distinguish similar Web services (Huang and Lin 2013). More and more researchers within this field pay close attention to QoS issues, and the methods for Web service selection and recommendation based on QoS are very popular now.

However, there has been a long controversy about the real-time acquisition of QoS attributes (Shao, Zhang *et al.* 2007; Zheng, Ma *et al.* 2009). On the one hand, within a short time it is almost impossible for a service requestor to invoke all of the candidate Web services in question with limited computing resources. On the other hand, the values of QoS attributes are highly related to geographical location, service time and network condition, and they always change over time. For example, different users may obtain completely different QoS values when invoking the same Web service, and maybe the observed QoS values are smaller or larger than the corresponding value released by the provider of the Web service. Therefore, due to the underlying assumption that service consumers tend to obtain the best recommendations from those with similar QoS preferences or usage experiences to themselves, personalized QoS-aware Web service recommendation appears as an emerging technique to address the above issue.

As we know, CF (Collaborative Filtering), also known as social filtering or social information filtering, is the most popular technique in the field of personalized recommender systems. It aims to predict and identify the information (e.g., website, commodity, social networking service, etc.) a user might be interested in according to historical data, and to make recommendations on this basis. To the best of our knowledge, there are mainly two types of classic CF methods for Web service recommendation: memory-based and model-based approaches. Despite some criticism, the CF-based methods have been widely used in prior studies and in many commercial systems, and their feasibility and good performance have also been validated in terms of different data sets (Bobadilla, Ortega *et al.* 2013).

In general, the memory-based approaches can be divided into three categories: user-based (Breese, Heckerman *et al.* 1998; Jin, Chai *et al.* 2004), item-based (Deshpande, Karypis *et al.* 2004; Sarwar, Karypis *et al.* 2001), and hybrid approaches (Zheng, Ma *et al.* 2009; Zheng, Ma *et al.* 2011). The basic idea of this type of approaches is to conduct rating predictions based on historical QoS records after finding out similar users or items. Although they are easy to implement and are cost effective, there are several drawbacks with this type of approaches, such as the bias of human ratings and the relatively poor scalability with large-scale data sets.

On the other hand, the model-based approaches, such as latent factor models (LFMs), have gradually become a hot topic of Web service recommendation in both academia and industry because of the famous contest of Netflix Prize. In this model, users and Web services are mapped into the same latent space by decomposing the user-service QoS matrix into two low-dimensional matrices, and the rating of a given Web service is predicted by getting the inner product of the two matrices (Yu, Liu *et al.* 2014). The singular value decomposition (SVD) is one of the most frequently-used matrix factorization models (Hofmann, 2004). Generally speaking, this type of approaches can achieve better prediction performance and scalability with large-scale data sets and has a better ability to handle the sparsity, but their main disadvantages are in the expensive model building.

Considering the merits and demerits of memory-based and model-based approaches, a number of hybrid methods combining the two types of approaches have successively been proposed and applied to several application domains, e.g., image (Zhou, Cheung *et al.* 2010), movie (Pennock, Horvitz *et al.* 2000; Rashid, Lam *et al.* 2007), music (Yoshii, Goto *et al.* 2008), TV programs (Barragáns-Martíneza, Costa-Montenegroa *et al.* 2010), traffic (Alecsandru and Ishak 2004), commercial contest (Takács, Pilászy *et al.* 2008), etc. Inspired by that, a few researchers in the field of Services Computing attempted to perform personalized QoS-aware Web service recommendation using such type of hybrid approaches (Chen, Liu *et al.* 2010; Cao, Wu *et al.* 2013), and the experimental results show that the hybrid methods outperform those well-known CF algorithms with respect to prediction accuracy and scalability.

Overall, the main goal of rating-oriented (memory-based, model-based, or hybrid) CF methods is to obtain better approximations of the missing or unknown ratings in a given user-item matrix. However, high accuracy of rating prediction may not necessarily lead to a satisfactory recommendation result (Zheng and Lyu 2013), because in some application scenarios service requestors are more likely to follow with interest the rankings of target Web services rather than their detailed QoS values. That is, if more than one recommendation exists, how to recommend the top K appropriate Web services is actually a quality ranking issue; moreover, the quality ranking of recommended Web services might have a greater effect on user's choice. Therefore, personalized Web service recommendation based on QoS ranking prediction may be a potentially valuable research topic within this field (Zheng, Wu *et al.* 2013).

Unfortunately, to the best of our knowledge, few prior studies investigate the combination of rating-oriented and ranking-oriented approaches, and little is known about the feasibility of such a hybrid approach and about its advantages over other competing approaches. So, to address the issue, in this paper we propose a ranking-oriented hybrid approach that combines item-based and model-based CF methods to realize personalized Web service recommendation in terms of QoS values. In particular, in our method the similarity of two Web services is calculated in accordance with the rankings based on the ratings given to the two Web services by common users who have invoked both of them, and after identifying similar Web services, the method predicts the ratings of those Web services that have not been invoked at all using a hybrid CF model. Finally, the top K Web services are recommended to target users according to the overall ranking of all candidates in question. In brief, the primary contributions of this paper are described as follows:

(1) Considering the application requirement of Web services ranking, especially with respect to response time, we introduced ranking-oriented learning techniques to traditional item-based CF approaches, so as to find the most similar Web services in terms of the similarity based on ranking rather than rating. To validate the feasibility of our approach, we also improved the metric for the accuracy of a given method that returns the top K recommendations.

(2) We proposed a ranking-oriented hybrid method for personalized QoS-aware Web service recommendation, which integrated baseline estimate, item-based CF recommendation algorithm and latent factor model into a unified model. The experiment on a well-known data set indicates that our method outperforms other competing approaches in terms of the evaluation metric.

The remainder of this paper is organized as follows. Section RELATED WORK presents some typical prior studies related to the topic of this paper. In section QOS-AWARE WEB SERVICE RECOMMENDATION, we introduce the problem to be solved and the overall framework of our approach. In section RANKING-ORIENTED HYBRID APPROACH, a ranking-oriented hybrid CF approach combining item-based and model-based approaches is proposed to achieve better prediction performance. Section EXPERIMENTS AND PRIMARY RESULTS presents the experiments and some important findings obtained in our study. Finally, the conclusion and future work of this paper are summarized in section CONCLUSION AND FUTURE WORK.

## RELATED WORK

As the number of Web services available on the Internet increases quickly, service consumers pay more attention to QoS instead of functionality than before. QoS mainly consists of non-functional attributes such as response time, throughput, availability, etc. It has been widely used in service selection (Wang, Wang *et al.* 2013), service composition (Feng, Ngan *et al.* 2013), service recommendation (Cao, Wu *et al.* 2013; Jiang, Liu *et al.* 2011) and other popular topics in the field of Services Computing. In this section, we present the related work of QoS-aware Web service recommendation.

**Rating-Oriented Services Recommender**

The rating-oriented CF recommender is undoubtedly one of the most widely used approaches in the field of recommender systems, aiming at achieving better prediction accuracy of the missing QoS values for different service requestors. In general, it has two broad categories: memory-based and model-based approaches.

The memory-based CF approaches focus mainly on the similarity between users or items and can be classified as user-based approach (Breese, Heckerman *et al.* 1998; Jin, Chai *et al.* 2004), item-based approach (Deshpande, Karypis *et al.* 2004; Sarwar, Karypis *et al.* 2001) and hybrid approach (Zheng, Ma *et al.* 2009; Zheng, Ma *et al.* 2011). In 2007, Shao *et al.* introduced CF into Web service recommendation and proposed a classic user-based CF approach (Shao, Zhang *et al.* 2007). Subsequently, a series of user-based or item-based Web service recommendation methods were presented by many researchers within this field. In consideration of the advantages of user-based and item-based approaches, Zheng *et al.* (Zheng, Ma *et al.* 2009) proposed a new mixed model that integrated user-based and item-based approaches linearly by confidence weights, and the experimental results showed that the model and its improved version (Zheng, Ma *et al.* 2011) were able to achieve higher recommendation accuracy than those CF methods that belong to a single type. In recent literature of QoS-aware Web service recommendation, a few researchers attempted to incorporate the context including geographical location information (Tang, Jiang *et al.* 2012) and invocation time information (Zhang, Zheng *et al.* 2011a) into neighbor-based CF methods, and the leading advantages of these approaches with respect to recommendation performance were validated by large-scale experiments on real-world Web service QoS data sets.

Unlike simple and effective memory-based CF approaches, the model-based CF approaches introduce data mining, machine learning techniques to find patterns or train a prediction model based on training data. This type of approaches mainly includes clustering models (Xue, Lin *et al.* 2005), LFMs (Mnih and Salakhutdinov 2007), Bayesian networks (Singla and Richardson 2008), etc. Among these approaches, LFMs may be the most widely used one for Web service recommendation recently. LFMs have a more holistic goal to explain the interactions between users and Web services by analyzing the user-service matrix of historical QoS ratings, and matrix factorization (also known as matrix decomposition) techniques are a class of widely used LFMs. Zheng *et al.* (Zheng, Ma *et al.* 2013) proposed a collaborative Web service QoS prediction method via neighborhood integrated matrix factorization to predict those missing QoS values. The theoretical basis of the matrix factorization is that only a small part of significant factors affect QoS ratings in a given user-service matrix. Therefore, the objective of a LFM is to uncover latent factors that can explain observed ratings, and to classify the users and Web services in question by these factors. However, it should be pointed out that these classification factors are latent attributes which sometimes cannot be interpreted. Recently, Yu *et al.* (Yu, 2012; Yu, Zheng *et al.* 2013) presented a combination of the matrix factorization model and decision tree model for QoS prediction and utilized efficient iterative algorithms to solve the problem of QoS matrix completion, and similar study was also reported in literature (Yu, Liu *et al.* 2014). The experimental results showed that despite expensive model building, these approaches were more effective than the memory-based rivals, especially in the situation of data sparsity.

Besides, as mentioned above, a number of hybrid approaches, which combine memory-based and model-based approaches, have also been proven to be successful with respect to prediction performance in different application domains, such as music, movie and Web service (Chen, Liu *et al.* 2010; Cao, Wu *et al.* 2013). For more details of these rating-oriented CF methods, please refer to the recent surveys (Su and Khoshgoftaar 2009; Bobadilla, Ortega *et al.* 2013).

**Ranking-Oriented Services Recommender**

The rating-oriented CF approaches attempt to predict the vacant values in a given user-item matrix as accurately as possible, but in some real-world application scenarios, accurate rating predictions do not definitely lead to better recommendation performance (Zheng and Lyu 2013). For example, after the user $u$ invoked two Web services $s_i$ and $s_j$, the observed QoS values about response time (in seconds) are 0.4 and 0.5, respectively. Suppose that the QoS ratings of $s_i$ and $s_j$ (denoted by $\{q_i, q_j\}$) predicted by the recommendation models under discussion $M_1$ and $M_2$ are $\{0.3, 0.6\}$ and $\{0.5, 0.45\}$, respectively, it is clear that $M_2$ is better than $M_1$ in terms of root mean square error (RMSE). Therefore, the system will recommend $s_j$ to the users similar to $u$ according to the model $M_2$, which is obviously improper in practice since $s_i$ has a higher rank than $s_j$ with respect to response time.

Thus, the ranking-oriented recommender systems are more suitable for these application scenarios or requirements. The earlier study on the problem of learning how to order was conducted by Cohen *et al.* (Cohen, Schapire *et al.* 1997), and they proposed a greedy algorithm that was able to find a good approximation of the optimal ranking. Then, the related techniques and methods were introduced to the field of recommender systems. For example, to address the item ranking problem, Liu *et al.* (Liu and Yang 2008) proposed a ranking-based CF approach to movies recommendation, and the experimental result showed that their method outperformed traditional CF approaches significantly in terms of NDCG (Normalized Discounted Cumulative Gain). Yang *et al.* (Yang, Wei *et al.* 2009) also proposed a ranking-oriented CF method to solve the problem of the lack of user ratings, and their method achieved satisfactory effects on digital books recommendation based on users' access logs. Inspired by the topic models, Liu *et al.* (Liu, Chen *et al.* 2011) proposed an item-oriented model-based CF framework by user interest expansion via personalized ranking, which could address the problems of traditional CF approaches such as overspecialization and cold-start. According to matrix factorization models, Balakrishnan *et al.* (Balakrishnan and Chopra 2012) proposed a novel model that learned the features associated with the users and items for a ranking task, aiming at approximately optimizing NDCG for a given recommendation task. For more details of the ranking-oriented techniques for recommendation, please refer to the literature (Adomavicius and Kwon 2012).

In the field of Services Computing, as far as we know, only a few of researchers attempted to conduct Web service recommendation based on QoS ranking prediction recently. For example, Zheng *et al* (Zheng, Wu *et al.* 2013; Zheng and Lyu 2013) proposed a QoS-aware services ranking prediction framework based on the work mentioned above, and the superiority of the proposed methods to other related CF approaches was validated by the comprehensive experiments on real-world QoS data.

Different from the work mentioned above, in this paper we will propose a combination of rating-oriented and ranking-oriented CF approaches to perform personalized QoS-aware Web service recommendation, which not only meets the ranking demand on QoS values, but also takes advantage of the rating-oriented hybrid CF recommendation approaches.

# QOS-AWARE WEB SERVICE RECOMMENDATION

## Problem Definition

First of all, let us consider a toy example of QoS-aware Web service recommendation depicted in Figure 1. This bipartite graph $G = (U, S, E)$ includes two disjoint sets $U$ and $S$ which represent the sets of users and Web services, respectively. A weighted edge $e_{u,s}$ in the graph corresponds to the (historical) QoS value (e.g., response time in this example) of an invocation in the user-service QoS matrix (see Table 1). The primary aim of rating-oriented CF approaches is to effectively predict the weights of potential invocations, i.e. the blank spaces in Figure 2. Since accurate QoS rating prediction may not lead to satisfactory QoS ranking prediction (Zheng and Lyu 2013), for the ranking-oriented hybrid CF approach in this paper, one of our primary tasks is to predict the ranking of the top K Web services with respect to QoS directly.

Then, we formally define the problem of Web services QoS ranking prediction as follows.

$U = \{u_1, u_2, …, u_m\}$ is a set of users, where $m$ is the total number of users in the system.

$S = \{s_1, s_2, …, s_n\}$ is a set of Web services, where $n$ is the total number of Web service in the system.

$Q = (q_{u,s})_{m \times n}$ is a user-service matrix of historical QoS values, where each entry $q_{u,s}$ represents the QoS value of the Web service $s$ observed by the user $u$. If there is no past experience rated by the user $u$ on the Web service $s$, $q_{u,s} \in \emptyset$.

$\text{RecList}_u = \{s_i^{\text{Rec}} \mid s_i^{\text{Rec}} \in \text{Rec}_K^u(S), 1 \leq i \leq K\}$ is the list of recommendations to the target user $u$, where $\text{Rec}_K^u(S)$ is a partially ordered set of the top K Web services sorted in ascending or descending order by the existing and predicted QoS ratings. Besides, we also define $\text{IRecList}_u$ as the ideal list of recommended Web services sorted in the same order as $\text{RecList}_u$ by QoS ratings in test data set in which the user-service matrix is sound and intact.

Hence, the problem of this paper is how to make RecList approximate IRecList (for all target users) as closely as possible, which can be formally defined as $\max \sum_u sim'(\text{RecList}_u, \text{IRecList}_u)$ ($K \geq 1$). Note that $sim'()$ is a function that calculates the similarity of two lists of Web services ($sim'() \in [0, 1]$).

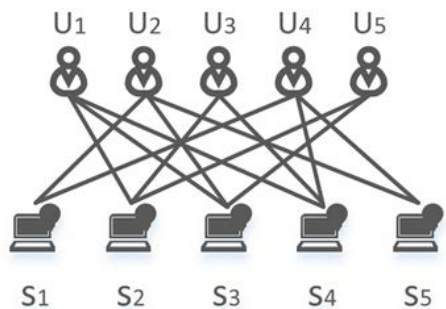

| | $s_1$ | $s_2$ | $s_3$ | $s_4$ | $s_5$ |
|---|---|---|---|---|---|
| $u_1$ | | 0.34 | 0.42 | 0.42 | |
| $u_2$ | 0.23 | | 0.43 | | 0.49 |
| $u_3$ | | 0.50 | | 0.33 | |
| $u_4$ | 0.38 | | | 0.24 | 0.45 |
| $u_5$ | | 0.48 | 0.51 | | |

Figure 1. User-Service Interaction Diagram       Figure 2. User-Service QoS Matrix

## Overview of our approach

As shown in Figure 3, our approach to personalized QoS-aware Web service recommendation has four main steps, and the details of each step will be described in the coming section.

First, for a given user-service matrix, the similarity between two Web services is measured by observing the rankings (rather than the ratings) of all the users who have rated both the items (i.e. Web services). Second, after identifying similar Web services according to the values of item similarity, we use only the Top-k similar neighbors to perform QoS rating predictions. Third, those missing values in the user-service matrix are predicted by a hybrid model that combines item-based CF algorithm, latent factor model and baseline estimate. Finally, our approach returns the top K Web services in terms of the overall ranking of (existing and predicted) QoS ratings to the target user.

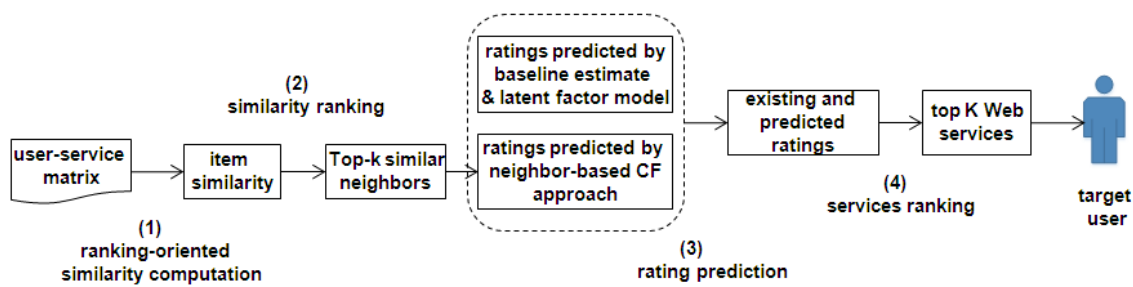

Figure 3. Overview of our approach

# RANKING-ORIENTED HYBRID APPROACH

## Ranking-Oriented Similarity Computation

Because the neighbor-based approaches are intuitive and relatively easy to implement, they have been deemed as the most popular CF method, including user-based and item-based approaches. Actually, the item-based approaches are more favorable for better scalability and improved accuracy in many cases (Bell and Koren 2007; Sarwar, Karypis *et al.* 2001; Takács, Pilászy *et al.* 2007). Therefore, in this paper our focus is on the item-based CF recommendation model. In general, the traditional rating-oriented approaches compute the similarity between items in terms of Pearson Correlation Coefficient (PCC), which measures the tendency of the two Web services in question to share the similar historical QoS records observed (or rated) by users. To address the shortcomings of rating-oriented CF approaches, in this paper we used the Kendall Rank Correlation Coefficient (KRCC) (Marden, 1996) to measure the item similarity between two rankings on the same set of common users' invocations, which can be defined as follows:

$$sim(s_i, s_j) = 1 - \frac{4 \times \sum_{u,v \in U_{s_i} \cap U_{s_j}} g((q_{u,s_i} - q_{u,s_j})(q_{v,s_i} - q_{v,s_j}))}{|U_{s_i} \cap U_{s_j}| \times (|U_{s_i} \cap U_{s_j}| - 1)}, \quad (1)$$

where $U_{s_i} \cap U_{s_j}$ is the set of users who commonly invoked the Web services $s_i$ and $s_j$ and $g(x)$ is an indicator function (as defined below).

$$g(x) = \begin{cases} 1 & x < 0 \\ 0 & x \geq 0 \end{cases}. \tag{2}$$

According to the definition, the KRCC similarity of two rankings is inversely proportional to the number of discordant pairs between the two rankings, and its value ranges from -1 to 1. Here, if $(q_{u,s_i} - q_{u,s_j})(q_{v,s_i} - q_{v,s_j}) < 0$, that's a discordant pair. When the two rankings are completely identical, the value of the similarity equals 1. Conversely, its value is equal to -1. It is worthy to note that the common invocations of the Web services $s_i$ and $s_j$ conducted by users has to be at least 2 ($|U_{s_i} \cap U_{s_j}| \geq 2$) since the metric KRCC compares user pairs. In other words, all the two rankings to be compared contain at least two elements.

**Finding Similar Items**

After obtaining the similarity values between different Web services, the similar items among all samples can be identified. It is very important to select the most similar neighbors to make accurate recommendations, because the neighbors with a low similarity score may decrease the prediction accuracy greatly. As we know, the traditional Top K algorithm is often used to solve this problem by choosing the Top-k most similar neighbors. In this paper, the set of the Top-k closest neighbors based on the item similarity (see Equation (1)) excludes the neighbors whose similarity values are equal to or smaller than 0. The set of the Top-k most similar Web services for the Web service $s_i$ ($S^k(s_i)$) is identified by

$$S^k(s_i) = \{s_j \mid s_j \in T(s_i), sim(s_i, s_j) > 0, i \neq j\}, \tag{3}$$

where $T(s_i)$ is a set of the Top-k most similar Web services to the target Web service $s_i$.

**QoS Rating Prediction**

In this paper, the item-based CF recommendation model uses the most similar Web services set (see Equation (3)) to predict the missing values in a given user-service QoS matrix in terms of the following basic equation.

$$\hat{q}_{u,s_i} = b_{u,s_i} + \sum_{s_j \in S^k(s_i)} \frac{sim(s_i, s_j)}{\sum_{s_j \in S^k(s_i)} sim(s_i, s_j)} \cdot (q_{u,s_j} - b_{u,s_j}), \tag{4}$$

where $b_{u,s_i}$ is the baseline estimate of the Web service $s_i$ invoked by the user $u$.

This model can be adjusted with the interaction effect between users and items by means of the baseline estimate, so that it has an ability to achieve better prediction accuracy. In the following subsections, we will introduce this model and its improved version in detail.

**Baseline Estimate**

The typical CF approaches analyze the interaction effect between users and Web services, and they tend to predict higher ratings for the users whose historical QoS records are, on average, higher than others. In this paper, the influence of users and the quality of Web services on QoS rating (without consideration of the effect of their interactions) is measured by using the baseline estimate. A baseline estimate $b_{u,s_i}$ for an unknown rating $\hat{q}_{u,s_i}$ is defined as follows:

$$b_{u,s_i} = \mu + b_u + b_{s_i}, \qquad (5)$$

where $\mu$ is the average rating, and the parameters $b_u$ and $b_{s_i}$ indicate the observed deviations of the user $u$ and the Web service $s_i$ from the average rating, respectively. For example, suppose that we want to calculate the baseline estimate for the response time (unit: second) of the Web service $s_i$ invoked by the user $u$. If the average value of the global response time in the user-service matrix is 0.3, the response time of the Web service is decreased by 0.1 compared with the average (due to better network environment), and the poor computing facility of the user makes the response time 0.05 higher than the average, the baseline estimate for the rating of the Web service in question will be calculated as $b_{u,s_i} = \mu + b_u + b_{s_i} = 0.3 + 0.05 + (-0.1) = 0.25$.

It should be noted that the goal of the baseline estimate is not to simply calculate the averages of users and items. In order to estimate the parameters $b_u$ and $b_{s_i}$, one can solve the least squares problem with the stochastic parallel gradient descent algorithm (see Line 5 and Line 6 of Algorithm 1 in Figure 4).

**Latent Factor Model**

As we know, LFMs can explore the latent attributes which cause the QoS ratings with a more holistic goal, and the classic methods include Probabilistic Latent Semantic Analysis (PLSA) (Hofmann, 2004), neural networks (Salakhutdinov, Mnih *et al.* 2007) and Latent Dirichlet Allocation (LDA) (Blei, Ng *et al.* 2003). In this paper, we employed SVD to factorize the user-service QoS rating matrix. SVD was mainly used in the field of natural language processing (Gorrell and Webb 2005) until the competition of Netflix Prize. Since then, it has been introduced to CF approaches (Funk, 2006) and has gained much attention due to its attractive prediction accuracy.

SVD matrix factorization models explore a latent description for each individual and factorize the matrix under discussion into two low-dimensional matrices. From the perspective of matrix factorization, a given user-service QoS matrix to be predicted can be factorized into two low-dimensional matrices (as defined below).

$$\hat{Q} = P^T W, \qquad (6)$$

where $P = P_{F \times m} = (p_{f,u})$ and $W = W_{F \times n} = (w_{f,s_i})$ are two sub-matrices, $F$ is the number of user-defined dimensions, and $m$ and $n$ are the total numbers of users and Web services in the system, respectively. The QoS rating prediction for the Web service $s_i$ invoked by the user $u$ is then defined as follows:

$$\hat{q}_{u,s_i} = b_{u,s_i} + \sum_f p_{u,f} w_{f,s_i}. \qquad (7)$$

Although SVD models can achieve more accurate rating prediction through analyzing the matrix in question, the user-service QoS rating matrices available are usually sparse, possibly resulting in the problem of over-fitting when directly training a model based on a small number of existing QoS records. So, the regularization term is added to avoid the over-fitting problem. In order to estimate the parameters in this model, one can solve the least squares problem described as follows:

$$C(\mathbf{p}_u, \mathbf{w}_{s_i}, \mathbf{b}_u, \mathbf{b}_{s_i}) = \sum_{(u,s_i) \in T} \left[ (q_{u,s_i} - b_{u,s_i} - \sum_f p_{u,f} w_{f,s_i})^2 + \lambda_1 (\| \mathbf{p}_u \|^2 + \| \mathbf{w}_{s_i} \|^2 + \| \mathbf{b}_u \|^2 + \| \mathbf{b}_{s_i} \|^2) \right], \quad (8)$$

where $T$ represents a training set and $\lambda_1$ is a user-defined regularization parameter. The stochastic parallel gradient descent algorithm can also be applied to obtaining the recursion formulas of the matrices $P$ and $W$, similar to those indicated in Algorithm 1 (see Line 9 and Line 10 in Figure 4). After that, SVD can be easily implemented by MATLAB programming.

**Hybrid Recommendation Model**

As mentioned before, memory-based and model-based CF approaches can complement each other very well to analyze user-service QoS (historical ratings) matrices. Accordingly, in this subsection we integrate the two types of methods into our hybrid recommendation model named 2RHyRec, which is defined based on Formula (4) and Formula (7) as

$$\hat{q}_{u,s_i} = b_{u,s_i} + \beta \cdot \sum_{s_j \in S^k(s_i)} \frac{sim(s_i, s_j)}{\sum_{s_j \in S^k(s_i)} sim(s_i, s_j)} \cdot (q_{u,s_j} - b_{u,s_j}) + (1-\beta) \cdot \sum_f p_{u,f} w_{f,s_i}. \quad (9)$$

According to Formula (9), the 2RHyRec model consists of three types of recommendation approaches. First, the baseline estimate, $\mu + b_u + b_{s_i}$, provides the general features of the target user and Web service without considering the effect of any interactions involved. Second, the neighbor-based model, $\sum_{s_j \in S^k(s_i)} \frac{sim(s_i, s_j)}{\sum_{s_j \in S^k(s_i)} sim(s_i, s_j)} \cdot (q_{u,s_j} - b_{u,s_j})$, gives a fine correlation/association analysis of local neighbors based on the KRCC ranking similarity between Web services. Third, the latent factor model, $\sum_f p_{u,f} w_{f,s_i}$, describes the interactions between users and Web services from a global perspective. In the hybrid model, the customization parameter $\beta$ is in the interval of [0, 1] for model weight adjustment that can adapt to different research questions.

Besides, the parameters of our model are determined by minimizing prediction error on training data. As we know, there are several learning techniques to achieve the above objective. Thus, the model parameters can be learned by solving the regularized least squares problem associated with the following objective function:

$$\min_{\mathbf{b}_*, \mathbf{p}_*, \mathbf{w}_*} \sum_{(u,s_i) \in T} \left[ (q_{u,s_i} - \hat{q}_{u,s_i})^2 + \lambda \cdot (\| \mathbf{b}_u \|^2 + \| \mathbf{b}_{s_i} \|^2 + \| \mathbf{p}_u \|^2 + \| \mathbf{w}_{s_i} \|^2) \right], \quad (10)$$

where the first term $\sum_{(u,s_i) \in T} (q_{u,s_i} - \hat{q}_{u,s_i})^2$ is used to find the most suitable vectors $\mathbf{b}_*$, $\mathbf{p}_*$ and $\mathbf{w}_*$ to approximate the training data as closely as possible, and the regularization term

$\lambda \cdot (\|\mathbf{b}_u\|^2 + \|\mathbf{b}_{s_i}\|^2 + \|\mathbf{p}_u\|^2 + \|\mathbf{w}_{s_i}\|^2)$ is utilized to avoid over-fitting by penalizing the magnitudes of the parameters. Note that $\lambda$ is a customization parameter to measure the weight of parameters in penalization.

In this paper, Formula (10) (i.e. the loss function) was solved using stochastic gradient descent technique, and the corresponding algorithm of solving the model parameters is shown in Figure 4 (demonstrated in pseudo-code format). First, the algorithm took the corresponding partial derivative of the objective function with respect to each parameter in question, and set the partial derivatives equal to zero simultaneously to find the steepest descent direction. Second, the algorithm then optimized the parameters with iterative methods on training data. Third, in each iteration of the learning process, the algorithm reduced the learning rate ($\alpha$) representing the rate of gradient descent along with the decrease of the loss function. Finally, as the number of iterations increased, if the prediction error of this algorithm gradually decreased until a certain value was reached, the learning process was then terminated. Fortunately, the convergence of stochastic gradient descent has been analyzed and validated using the theories of convex minimization and stochastic approximation.

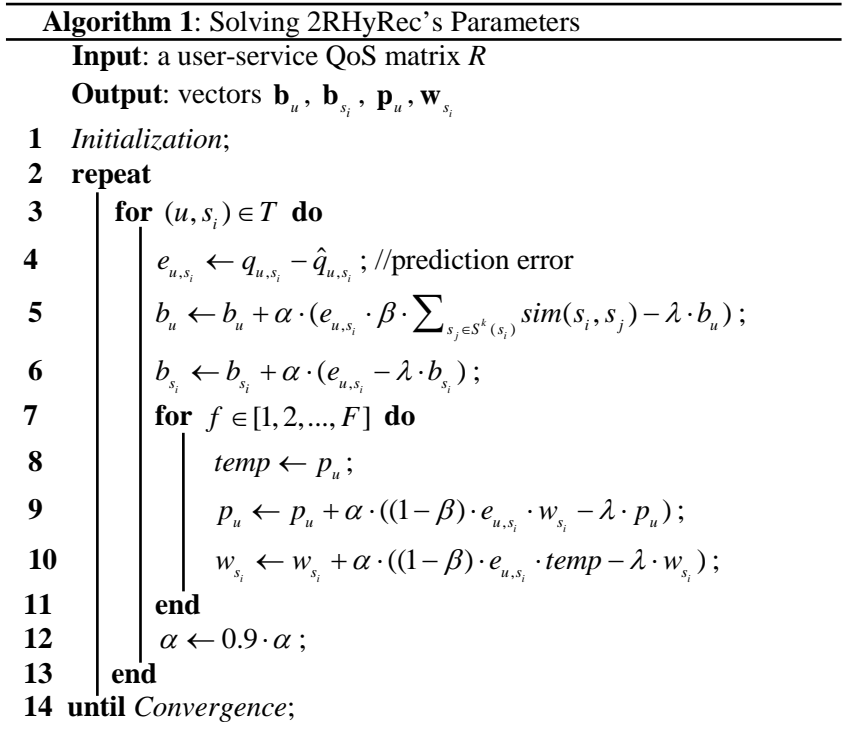

**Algorithm 1**: Solving 2RHyRec's Parameters
**Input**: a user-service QoS matrix $R$
**Output**: vectors $\mathbf{b}_u$, $\mathbf{b}_{s_i}$, $\mathbf{p}_u$, $\mathbf{w}_{s_i}$
1 *Initialization*;
2 **repeat**
3   **for** $(u, s_i) \in T$ **do**
4     $e_{u,s_i} \leftarrow q_{u,s_i} - \hat{q}_{u,s_i}$; //prediction error
5     $b_u \leftarrow b_u + \alpha \cdot (e_{u,s_i} \cdot \beta \cdot \sum_{s_j \in S^k(s_i)} sim(s_i, s_j) - \lambda \cdot b_u)$;
6     $b_{s_i} \leftarrow b_{s_i} + \alpha \cdot (e_{u,s_i} - \lambda \cdot b_{s_i})$;
7     **for** $f \in [1, 2, ..., F]$ **do**
8       $temp \leftarrow p_u$;
9       $p_u \leftarrow p_u + \alpha \cdot ((1-\beta) \cdot e_{u,s_i} \cdot w_{s_i} - \lambda \cdot p_u)$;
10       $w_{s_i} \leftarrow w_{s_i} + \alpha \cdot ((1-\beta) \cdot e_{u,s_i} \cdot temp - \lambda \cdot w_{s_i})$;
11     **end**
12     $\alpha \leftarrow 0.9 \cdot \alpha$;
13   **end**
14 **until** *Convergence*;

Figure 4. The algorithm of solving the parameters of 2RHyRec

If a training set contains $m$ common users who have rated both the Web services $s_i$ and $s_j$, the time complexity of calculating $sim(s_i, s_j)$ in terms of KRCC is $O(m^2)$, because there are at most $m(m-1)/2$ user pairs for these common users. To find similar Web services for the Web service $s_i$, we have to calculate the similarities between $s_i$ and all the $n$ training samples ($n \geq m$), namely, there are $n$ times of similarity computations. Therefore, for the training set, the total time complexity of similarity computation based on KRCC is $O(m^2n^2)$. On the other hand, since the training set in this paper is an $n \times n$ matrix, the time complexity of SVD for the matrix is $O(Fn^3)$. According to the above analysis, the computational complexity of Algorithm 1 at every iteration is $O(\max\{m^2n^2, Fn^3\})$.

### Ranking candidate Web services

After obtaining the predicted values for those missing elements in the user-service QoS rating matrix, personalized Web service recommendation can be easily performed based on the complete matrix. According to the target user's non-functional requirements or preference on QoS rating, all candidate Web services are sorted in a certain order. For example, the recommender arranges the values of candidate Web services in ascending order if the target user focuses on response time, while it will return the results in descending order when considering availability. Eventually, the top K Web services in the sorted list with respect to QoS rating are recommended to the target user.

## EXPERIMENTS AND PRIMARY RESULTS

### Research Questions

Since QoS rating prediction is a core component of QoS-aware Web service recommendation, in this paper we used the prediction performance of our approach to measure the quality of recommendations. To assess and evaluate our approach, the experiments we designed were conducted to answer the following research questions:

(1) Does our approach outperform other competing recommendation methods?

(2) How do the user-defined parameters *topK* (the number of recommended Web services), $\beta$, $F$ and Top-$k$ (the number of similar Web services, see Formula (3)) affect the prediction accuracy of our approach, respectively?

### Data Set Description

In this paper, we used a public data set of real-world Web services introduced in (Zheng, Zhang *et al.* 2010), which contains over one and a half millions QoS records from 339 users and 5825 Web services distributed all over the world. The density of the user-service matrix in this data set used for evaluation is 94.9%. For more details of this data set, please refer to the literature (Zheng, Zhang *et al.* 2010; Zhang, Zheng *et al.* 2011b). In our experiments, as with those prior studies, we first obtained a user-service QoS matrix (100×100 or 150×150) from the data set, where each entry in such a matrix is a vector including values of different QoS properties. We then randomly extracted a sub-matrix within the QoS matrix with a certain density (from 10% to 30%) as training data, and the remainder of the QoS matrix was used as test data to validate our approach. For example, Table 1 shows an example of a 100×100 user-service QoS matrix with respect to response time (unit: second).

Table 1. User-service QoS matrix with respect to response time

|  | $s_1$ | $s_2$ | ... | $s_{100}$ |
|---|---|---|---|---|
| $u_1$ | 5.982 | 0.228 | ... | 0.237 |
| $u_2$ | 2.13 | 0.262 | ... | 0.273 |
| ... | ... | ... | ... | ... |
| $u_{100}$ | 0.854 | 0.366 | ... | 0.376 |

**Evaluation Metric**

Because traditional rating-oriented recommendation approaches aim to predict QoS values as accurately as possible, the concept of deviation is often applied to measuring the prediction accuracy of the method in question. As we know, the two widely used evaluation metrics for rating-oriented CF approaches are mean absolute error (MAE) and root mean square error (RMSE). Generally speaking, the smaller values of MAE and RMSE indicate better prediction performance. Unlike those rating-oriented recommendation methods, in this paper we introduced Normalized Discounted Cumulative Gain (NDCG) to measure the quality of Web services ranking. NDCG was first used in the field of information retrieval (Järvelin and Kekäläinen 2002), and it is more suitable for evaluating ranking results compared with MAE and RMSE.

The original DCG-$k$ for a ranking of the top K recommended Web services is defined as follows:

$$\text{DCG-}k = rel_1 + \sum_{i=2}^{l} \frac{rel_i}{\log_2 i}, \qquad (11)$$

where $rel_i$ is the QoS rating of the $i^{th}$ Web service in the ranking. According to the definition, there exists a negative correlation between DCG-$k$ and the position $i$, which is determined by the decay curve of $\log_2 i$. However, in most cases, DCG-$k$ cannot be applied directly to ranking results since different queries from users may have ranking results of different sizes. So, NDCG-$k$ is proposed to meet the need for the comparison of different queries, and it is defined as follows:

$$\text{NDCG-}k = \frac{\text{DCG-}k}{\text{IDCG-}k}, \qquad (12)$$

where IDCG-$k$ is the ideal value of DCG-$k$ computed based on test data (used as ground truth).

The value of NDCG-$k$ is in the interval of [0, 1], with values closer to 1 indicating better ranking prediction, because the result is a close approximation to the ideal ranking. Considering the effectiveness of NDCG-$k$ in measuring ranking quality, in this paper it was also utilized as an evaluation metric to measure the ranking quality of recommended Web services.

**Design of Experiments**

To answer the first research question, we compared the 2RHyRec with other eight competing methods with respect to prediction performance in terms of the evaluation metric. The first six methods belong to the class of rating-oriented CF approaches, while the last two methods fall within the scope of ranking-oriented CF methods. The eight methods under discussion are described as follows.

- UMEAN: the user-based mean rating method for blank QoS rating values;
- IMEAN: the item-based mean rating method for blank QoS rating values;
- UPCC: the user-based collaborative filtering method using Pearson Correlation Coefficient (PCC) to measure the similarities between users (Breese, Heckerman *et al*. 1998);
- IPCC: the item-based collaborative filtering method with Pearson Correlation Coefficient (PCC) to measure the similarities between items (Resnick, Iacovou *et al*. 1994);
- WSRec: the hybrid model composed of UPCC and IPCC with confidence weight (Zheng, Ma *et al*. 2009);

- BiasSVD: the latent factor model using singular value decomposition with user and item bias (Paterek, 2007);
- GM: the greedy method for learning to order items (Cohen, Schapire *et al*. 1997);
- CloudRank2: the cloud service ranking method with confidence levels of different preference values (where the similarities between users are measure by KRCC) (Zheng, Wu *et al*. 2013).

To answer the second research question, we investigated the impact of each parameter in question on the value of NDCG-*k* with the method of multi-parameter adjustment control. That is, the other parameters of our approach were set to their own optimal values in advance, and we then varied the value of the parameter under discussion with a given step value to observe the change in the value of NDCG-*k*.

**Performance Comparison**

To compare the prediction performance of the above nine methods, the parameters of each method were set to their own optima for the data set by a computer program. Note that, for the 2RHyRec, $\lambda = 0.01$, $\alpha = 0.02$, $\beta = 0.6$, $F = 50$ and Top-$k$ = 20. In addition, to reduce the effect of randomly selecting training data, we carried out the experiment on each method in question for 10 times, and used the average of NDCG-*k* values as the final result.

For these methods under discussion, the prediction results with respect to response time are presented in Table 2, where *userNum* denotes the size of a given user-service QoS matrix and *matrixDes* represents the proportion of nonzero entries in the QoS matrix. For example, the value of *matrixDes* is equal to 10%, suggesting that we randomly select 10% of the QoS entries in the matrix. For each column in this table, the best performer among the nine rating-oriented and ranking-oriented approaches is highlighted in bold.

Table 2. Performance comparison of the methods in question with respect to response time

| Methods | userNum = 100 | | | | | | 150 | | | | | |
|---|---|---|---|---|---|---|---|---|---|---|---|---|
| | NDCG-10 | | | NDCG-20 | | | NDCG-10 | | | NDCG-20 | | |
| | matrixDes = 10% | 20% | 30% | 10% | 20% | 30% | 10% | 20% | 30% | 10% | 20% | 30% |
| UMEAN | 0.2973 | 0.3285 | 0.4257 | 0.3186 | 0.3507 | 0.4360 | 0.2786 | 0.2942 | 0.4106 | 0.3617 | 0.4584 | 0.4760 |
| IMEAN | 0.3291 | 0.3896 | 0.4783 | 0.3874 | 0.4460 | 0.4465 | 0.3593 | 0.4226 | 0.5084 | 0.5366 | 0.6146 | 0.6636 |
| UPCC | 0.3452 | 0.4508 | 0.5997 | 0.4060 | 0.5807 | 0.6497 | 0.3442 | 0.4761 | 0.5666 | 0.6162 | 0.5374 | 0.7138 |
| IPCC | 0.3510 | 0.4098 | 0.5630 | 0.4991 | 0.6126 | **0.7273** | 0.4158 | 0.4457 | 0.5953 | 0.6068 | 0.6297 | 0.7489 |
| WSRec | 0.3956 | 0.4781 | 0.5770 | 0.5472 | 0.6285 | 0.6644 | 0.4321 | 0.4933 | 0.5977 | 0.6228 | 0.6301 | 0.7791 |
| BiasSVD | 0.3747 | 0.4401 | 0.4547 | 0.4966 | 0.6134 | 0.6244 | 0.3684 | 0.4917 | 0.5241 | 0.5636 | 0.5836 | 0.7285 |
| GM | 0.4662 | 0.5067 | 0.5953 | 0.5656 | 0.6379 | 0.6885 | 0.4858 | 0.5251 | 0.6370 | 0.6505 | 0.6260 | 0.7726 |
| CloudRank2 | 0.4742 | 0.5319 | 0.6288 | 0.5803 | 0.6403 | 0.6926 | 0.5079 | 0.5292 | **0.6451** | 0.6603 | 0.7289 | 0.7885 |
| 2RHyRec | **0.5092** | **0.5950** | **0.6383** | **0.6166** | **0.6791** | 0.7211 | **0.5121** | **0.5497** | 0.6431 | **0.6840** | **0.7415** | **0.7893** |

The experimental results of Table 2 show that:

(1) Among all the nine methods, our approach can achieve the best prediction performance in terms of NDCG-*k* in most cases (10/12), indicating that it works better for personalized QoS-aware Web service recommendation on the whole, since its main goal is to address the ranking problem by making full use of the advantages of neighbor-based CF approaches and latent factor models.

(2) Considering the application scenario (or user requirements), the three ranking-oriented methods (viz. GM, CloudRank2 and 2RHyRec) are, in general, better than the six rating-oriented approaches with respect to prediction accuracy, especially when selecting a very small portion (*matrixDes* = 10%) of the 100×100 or 150×150 user-service QoS matrix as training data.

(3) For each method under discussion, its prediction accuracy is improved with an increase of the matrix density (from 10% to 30%) in the case of the same matrix size and evaluation metric, implying that the quantity of training data seems to contribute to higher prediction accuracy.

(4) In the experiment, the metric NDCG-*k* indicates that the ranking accuracy of the top K Web services is investigated. For the matrices with different sizes, each method in question achieves better prediction accuracy in terms of NDCG-20, which suggests that an appropriate value of the parameter *topK* may have a great effect on prediction performance.

(5) When we compared IPCC with UPCC, the former outperformed the latter under most experimental settings. The interesting observation indicates that the relationships among Web services in the training set are more useful in personalized QoS-aware Web service recommendation. That is the main reason why we integrated IPCC into our approach.

## Influence Analysis of Model Parameters

### Impact of *topK*

As mentioned earlier, the parameter *topK* determines the number of Web services recommended to the target user. To investigate the impact of *topK* on the accuracy of the 2RHyRec model, the values of other parameters *userNum*, *F*, Top-*k* and *β* were set to 100, 50, 20 and 0.6, respectively; moreover, the value of *topK* was changed from 5 to 45 with a step value of 5. The impact of *topK* on the prediction performance of our approach (with respect to response time) is presented in Figure 5, where the X-axis indicates the value of *topK* and the Y-axis represents the value of the evaluation metric.

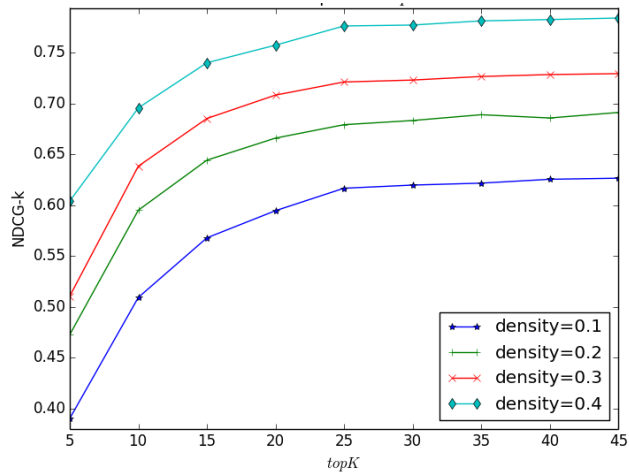

Figure 5. Impact of *topK*

It is apparent from Figure 5 that the four curves (representing four different matrix densities) exhibit a very similar trend in accuracy with the increasing of the value of *topK*. When the value of *topK* is relatively small, the accuracy of our approach can be obviously improved with the increasing of the parameter's value, especially when its value is increased from 5 to 10. On the contrary, when the parameter has a value greater than or equal to 25, the impact on the evaluation metric NDCG-*k* becomes less marked. It is worthy to note that we also find similar results when *userNum* = 150, *F* =50, Top-*k* = 20, and $\beta$ = 0.6, implying that the size of the input matrix has little influence on the results, but not all the information could be displayed here due to space limitations.

**Impact of $\beta$**

The item-based CF approaches and latent factor models have been widely used in the field of personalized Web service recommendation, and each type of methods has different advantages in analyzing a given user-service QoS matrix from diverse perspectives. As shown in Formula (9), $\beta$ is a customization parameter which determines the confidence weights of the two types of methods in our model. If the parameter equals 0, the 2RHyRec model is equivalent to a latent factor model, while it degenerates into a single neighbor-based model when $\beta$ = 1.

To investigate the impact of this parameter on our model's accuracy, the values of other parameters *userNum*, *F*, *topK* and Top-*k* were set to 100, 50, 20 and 20, respectively, and the value of $\beta$ was increased from 0.1 to 0.9 with a step value of 0.1. Note that, the value of *topK* was set to 20 because a value of 20 is a good compromise between accuracy and the number of recommended Web services in this paper.

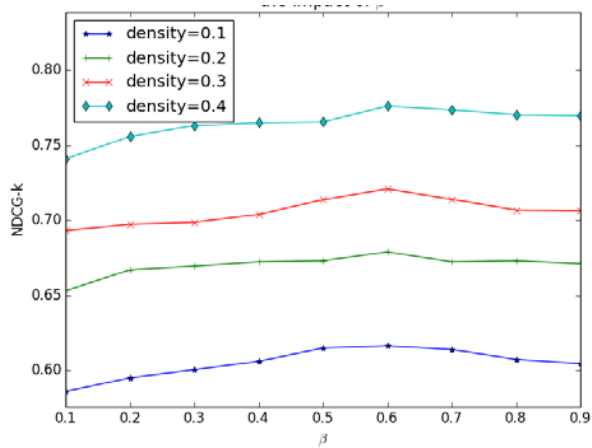

Figure 6. Impact of $\beta$

The analysis results are shown in Figure 6, where the X-axis and Y-axis represent the values of $\beta$ and NDCG-20, respectively. Interestingly, the four curves indicating four different matrix densities exhibit similar trends in accuracy. That is, the ascent part of each curve is presented in edge-up, and then slows down gradually after reaching peak ($\beta$ = 0.6). Therefore, such an optimum value of $\beta$ makes the most appropriate combination of the neighbor-based model and latent factor model. In addition, when we set *userNum* to 150, as the value of matrix density is increased from 10% to 40%, the optimum value of $\beta$ which obtains the maximum value of NDCG-20 is still 0.6, indicating that the weights of the two types of models are not influenced by the number of Web service QoS records.

**Impact of *F***

The dimensionality *F*, which determines the number of latent features, is used to factorize the user-service QoS matrix in the latent factor model of our approach. To investigate the impact of *F* on the prediction performance of our approach, we set the values of other parameters *userNum*, *β*, *topK* and Top-*k* to 100, 0.6, 20 and 20, respectively; moreover, the value of *F* was increased from 10 to 100 with a step value of 10. The analysis results are presented in Figure 7, where the X-axis and Y-axis indicate the values of *F* and NDCG-20, respectively.

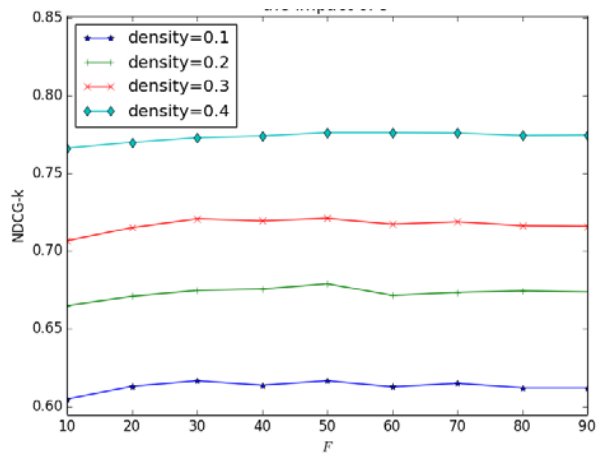

Figure 7. Impact of *F*

According to Figure 7, we can observe that under different conditions of matrix density, the performance of our approach changes slightly with the increasing of the value of *F*. Nonetheless, an appropriate value of *F* will lead to better performance. If *F* is too small, the low-dimensional sub-matrices are not good enough to describe those latent attributes, possibly resulting in our approach's relatively poor performance. On the other hand, although more latent dimensions make for more detailed description of the latent attributes, this always leads to over-fitting. Moreover, the larger the value of *F* is, the higher the time complexity of our approach becomes. For all the four curves in Figure 7, the recommendation accuracy gains the maximum when the value of *F* equals 50. In addition, when *userNum* is set to 150, we also find the same results as the matrix density is increased from 10% to 40%, suggesting that the size of a given user-service QoS matrix and matrix density have little influence on the optimal value of *F*.

**Impact of Top-*k***

The parameter Top-*k* determines the number of the most similar neighbors when finding similar Web services. To investigate the impact of Top-*k* on the accuracy of our approach, we set the values of other parameters *userNum*, *β*, *topK* and *F* to 100, 0.6, 20 and 50, respectively, and the value of Top-*k* was increased from 5 to 35 with a step value of 5. The analysis results are shown in Figure 8, where the X-axis and Y-axis represent the number of the most similar items and the value of NDCG-20, respectively.

Unlike the above parameters, there is no same or similar trend in accuracy among the four curves that represent different matrix densities. When the matrix density equals 10% and 30%, the results of Top-10 and Top-20 are very close to each other. When the matrix density is equal to 20%, Top-10 outperforms Top-20 in terms of NDCG-20, but the result is the opposite as the matrix density is increased to 40%. Generally speaking, the number 20 is an appropriate value of this parameter if you don't care about the computational cost, and this finding still holds when *userNum* is set to 150.

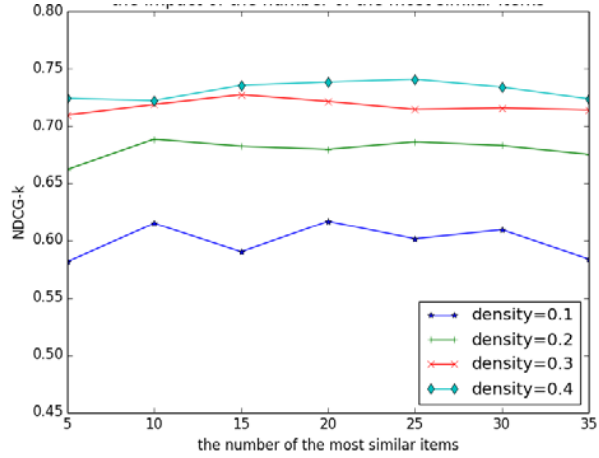

Figure 8. Impact of Top-*k*

## Discussion

**Convergence of the Algorithm**

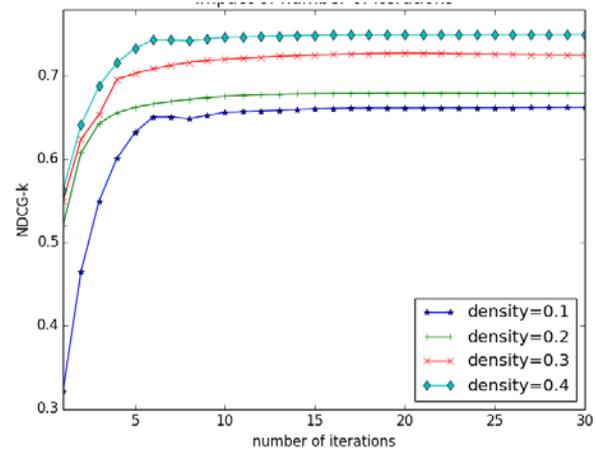

Figure 9. Convergence of Algorithm 1

Briefly, the implementation of Algorithm 1 uses an adaptive learning rate, that is, the learning rate $\alpha$ decreases with an appropriate rate and subjects to relatively mild assumptions. Prior studies on the convergence of such type of algorithms have proven that, stochastic gradient descent converges almost surely to a global minimum when the objective function is convex or pseudo-convex, and otherwise converges almost surely to a local minimum (Kiwiel, 2001; Kiwiel, 2003), which is actually a consequence of the Robbins-Siegmund theorem (Robbins and Siegmund 1971). Even so, it is still possible to fail in practice (for a given data set) due to too slow rate of convergence. In this paper, we evaluated the rate of convergence by means of empirical experiment rather than theoretical analysis. Note that the values of the parameters $\lambda$, $\alpha$, $\beta$, $F$, *topK*, Top-*k* and *userNum* were set to 0.01, 0.02, 0.6, 50 20, 20 and 100, respectively. The analysis results of Algorithm 1 with respect to convergence on the data set used in our experiments are shown in Figure 9, where the X-axis indicates the number of iterations and the Y-axis represents the value of NDCG-$k$. Each curve in Figure 9 presents the NDCG-$k$ value with a certain matrix density in each iteration step, and the result of the algorithm becomes steady after about ten iterations, indicating that the algorithm on this data set can converge efficiently.

**Threats to Validity**

So far, we have obtained several significant results to answer the two research questions, but potential threats to the validity of our work still remain.

Threats to *construct validity* are primarily related to the data set we used in this paper. It is collected by Zheng *et al.* (Zheng, Zhang *et al.* 2010) and includes only 339 users and 5,825 Web services. Although the data set has been validated and used in several prior studies, potential errors in the process of QoS records identification may exist. In addition, to ensure a comparison of different methods under the same conditions, in this paper we did not apply any data preprocessing techniques to the data set. On the other hand, considering the typical application scenario of Web services ranking, we used only a subset of the data set about response time.

Threats to *internal validity* are mainly related to the parameter settings of various methods in our study. To compare the best performance of each method under discussion, we sought the optimal value of each parameter in those methods with the method of multi-parameter adjustment control. Because some of the methods have several parameters, we had to change the value of the parameter under discussion with a given step value after the other parameters were assigned to preset values, but such an approach might lead to a problem that the value of the parameter may not be optimal. On the other hand, sometime, even if every parameter of a method is optimal, the method may not achieve the best performance.

Threats to *external validity* could be related to the generality of the results to other data sets used for personalized QoS-aware Web service recommendation. When we conducted our experiments, the QoS records in a given user-service QoS matrix were randomly selected according to matrix density to be training data. Although each experiment were repeated ten times, it is possible that we accidentally selected the data that has better or worse prediction accuracy than the average prediction performance by using the total data in the data set. Therefore, we need to validate the generality of our findings on more large-scale data sets.

# CONCLUSION AND FUTURE WORK

In this paper, we proposed a QoS-aware ranking-oriented hybrid Web service recommendation approach (named 2RHyRec), aiming at the ranking issue in predicting the missing QoS values in a given data set. By combining the advantages of the neighbor-based CF approaches and latent factor models in analyzing the user-service QoS matrix from different perspectives, our approach can obtain a higher accuracy rate than other competing approaches in terms of the metric NDCG-$k$. Meanwhile, the interpretability of our approach has been greatly improved by the introduction of an item-based model. Experimental results on a data set composed of real-world Web services show that our approach outstrips the existing typical rating-oriented and ranking-oriented methods with respect to accuracy. In addition, we also discussed the impacts of the user-defined parameters in our approach on the performance of the hybrid approach.

For the future work, we plan to learn more about the latent characteristics of the historical QoS data in other large-scale data sets, and to conduct more experiments to improve the prediction accuracy by integrating the context-aware techniques into our approach.

# ACKNOWLEGMENT

This work is supported by the National Basic Research Program of China (No. 2014CB340401), the National Natural Science Foundation of China (Nos. 61272111and 61273216), the Youth Chenguang Project of Science and Technology of Wuhan City (No. 2014070404010232), and the open foundation of Hubei Provincial Key Laboratory of Intelligent Information Processing and Real-time Industrial System (No. znss2013B017).

## REFERENCES

O'reilly, T. (2007). What is Web 2.0: Design patterns and business models for the next generation of software, *Communications and Strategies*, 65(1), 17-37.

Shao, L., Zhang, J., Wei, Y., Zhao, J., Xie, B., Mei, H. (2007). Personalized QoS prediction for web services via collaborative filtering, *Proceedings of the 14$^{th}$ IEEE International Conference on Web Services (ICWS 2007)*, Salt Lake City, Utah, USA, July 9-13, 2007, 439-446.

Huang, J., Lin, C. (2013). Agent-based green web service selection and dynamic speed scaling, *Proceedings of the 20$^{th}$ IEEE International Conference on Web Services (ICWS 2013)*, Santa Clara, CA, USA, June 28 - July 3, 2013, 91-98.

Zheng, Z., Ma, H., Lyu, M. R., King, I. (2009). Wsrec: A collaborative filtering based web service recommender system, *Proceedings of the 16$^{th}$ IEEE International Conference on Web Services (ICWS 2009)*, Los Angeles, CA, USA, July 6-10, 2009, 437-444.

Bobadilla, J., Ortega, F., Hernando, A., GutiéRrez, A. (2013). Recommender systems survey, *Knowledge-Based Systems*, 46, 109-132.

Breese, J. S., Heckerman, D., Kadie, C. (1998). Empirical analysis of predictive algorithms for collaborative filtering, *Proceedings of the 14$^{th}$ Conference on Uncertainty in Artificial Intelligence (UAI 1998)*, Madison, Wisconsin, USA, July 24-26, 1998, 43-52.

Jin, R., Chai, J. Y., Si, L. (2004). An automatic weighting scheme for collaborative filtering, *Proceedings of the 27$^{th}$ Annual International ACM SIGIR Conference on Research and Development in Information Retrieval (SIGIR 2004)*, Sheffield, UK, July 25-29, 2004, 337-344.

Deshpande, M., Karypis, G. (2004). Item-based top-n recommendation algorithms, *ACM Transactions on Information Systems*, 22(1), 143-177.

Sarwar, B. M., Karypis, G., Konstan, J. A., Riedl, J. (2001). Item-based collaborative filtering recommendation algorithms, *Proceedings of the 10$^{th}$ International Conference on World Wide Web (WWW 2001)*, Hong Kong, China, May 1-5, 2001, 285-295.

Zheng, Z., Ma, H., Lyu, M. R., King, I. (2011). Qos-aware web service recommendation by collaborative filtering, *IEEE Transactions on Services Computing*, 4(2), 140-152.

Yu, D., Liu, Y., Xu, Y., Yin, Y. (2014). Personalized QoS Prediction for Web Services Using Latent Factor Models, *Proceedings of the 11$^{th}$ IEEE International Conference on Services Computing (SCC 2014)*, Anchorage, AK, USA, June 27 - July 2, 2014, 107-114.

Hofmann, T. (2004). Latent semantic models for collaborative filtering, *ACM Transactions on Information Systems*, 22(1), 89-115.

Zhou, N., Cheung, W. K., Qiu, G., Xue, X. (2010). A Hybrid Probabilistic Model for Unified Collaborative and Content-Based Image Tagging, *IEEE Transactions on Pattern Analysis and Machine Intelligence*, 33(7), 1281-1294.

Pennock, D., Horvitz, E., Lawrence, S., Giles, C. L. (2000). Collaborative filtering by personality diagnosis: a hybrid memory- and model-based approach, *Proceedings of the 16$^{th}$ conference on Uncertainty in Artificial Intelligence (UAI 2000)*, Stanford, California, USA, June 30 - July 3, 2000, 473-480.

Rashid, A. M., Lam, S. K., LaPitz, A., Karypis, G., Riedl, J. (2007). Towards a Scalable kNN CF Algorithm: Exploring Effective Applications of Clustering, *Lecture Notes in Computer Science*, 4811, 147-166.

Yoshii, K., Goto, M., Komatani, K., Ogata, T., Okuno, H. (2008). An Efficient Hybrid Music Recommender System Using an Incrementally Trainable Probabilistic Generative Model, *IEEE Transactions on Audio, Speech, and Language Processing*, 16(2), 435 -447.

Barragáns-Martíneza, A., Costa-Montenegroa, E., Burguilloa, J., Rey-Lópezb, M., Mikic-Fontea, F., Peleteiroa, A. (2010). A hybrid content-based and item-based collaborative filtering approach to recommend TV programs enhanced with singular value decomposition, *Information Sciences*, 180(22), 4290–4311.

Alecsandru, C., Ishak, S. (2004). Hybrid Model-Based and Memory-Based Traffic Prediction System, *Transportation Research Record: Journal of the Transportation Research Board*, 1879, 59-70.

Takács, G., Pilászy, I., Németh, B., Tikk, D. (2008). Matrix factorization and neighbor based algorithms for the netflix prize problem, *Proceedings of the 2$^{nd}$ ACM conference on Recommender systems (RecSys 2008)*, Lausanne, Switzerland, October 23-25, 2008, 267-274.

Chen, X., Liu, X., Huang, Z., Sun, H. (2010). RegionKNN: A Scalable Hybrid Collaborative Filtering Algorithm for Personalized Web Service Recommendation, *Proceedings of the 17$^{th}$ IEEE International Conference on Web Services (ICWS 2010)*, Miami, Florida, USA, July 5-10, 2010, 9-16.

Cao, J., Wu, Z., Wang, Y., Zhuang, Y. (2013). Hybrid collaborative filtering algorithm for bidirectional Web service recommendation, *Knowledge and information systems*, 36(3), 607-627.

Zheng, Z., Wu, X., Zhang, Y., Lyu, M. R., Wang, J. (2013). QoS ranking prediction for cloud services, *IEEE Transactions on Parallel and Distributed Systems*, 24(6), 1213-1222.

Wang, X., Wang, Z., Xu, X. (2013). An Improved Artificial Bee Colony Approach to QoS-Aware Service Selection, *Proceedings of the 20$^{th}$ IEEE International Conference on Web Services (ICWS 2013)*, Santa Clara, CA, USA, June 28 - July 3, 2013, 395-402.

Feng, Y., Ngan, L. D., Kanagasabai, R. (2013). Dynamic Service Composition with Service-Dependent QoS Attributes, *Proceedings of the 20$^{th}$ IEEE International Conference on Web Services (ICWS 2013)*, Santa Clara, CA, USA, June 28 - July 3, 2013, 10-17.

Jiang, Y., Liu, J., Tang, M., Liu, X. (2011). An effective web service recommendation method based on personalized collaborative filtering, *Proceedings of the 18$^{th}$ IEEE International Conference on Web Services (ICWS 2011)*, Washington, DC, USA, July 4-9, 2011, 211-218.

Tang, M., Jiang, Y., Liu, J., Liu, X. (2012). Location-aware collaborative filtering for QoS-based service recommendation, *Proceedings of the 19$^{th}$ IEEE International Conference on Web Services (ICWS 2012)*, Honolulu, HI, USA, June 24-29, 2012, 202-209.

Zhang, Y., Zheng, Z., Lyu, M. R. (2011). WSPred: A time-aware personalized QoS prediction framework for Web services, *Proceedings of the 22$^{nd}$ IEEE International Symposium on Software Reliability Engineering (ISSRE 2011)*, Hiroshima, Japan, November 29 - December 2, 2011, 210-219.

Xue, G. R., Lin, C., Yang, Q., Xi, W., Zeng, H. J., Yu, Y., Chen, Z. (2005). Scalable collaborative filtering using cluster-based smoothing, *Proceedings of the 28$^{th}$ Annual International ACM SIGIR Conference on Research and Development in Information Retrieval (SIGIR 2005)*, Salvador, Brazil, August 15-19, 2005, 114-121.

Mnih, A., Salakhutdinov, R. (2007). Probabilistic matrix factorization, *Advances in neural information processing systems*, 20, 1257-1264.

Singla, P., Richardson, M. (2008). Yes, there is a correlation:-from social networks to personal behavior on the web, *Proceedings of the 17th International Conference on World Wide Web (WWW 2008)*, Beijing, China, April 21-25, 2008, 655-664.

Zheng, Z., Ma, H., Lyu, M. R., King, I. (2013). Collaborative web service qos prediction via neighborhood integrated matrix factorization, *IEEE Transactions on Services Computing*, 6(3), 289-299.

Yu, Q. (2012). Decision tree learning from incomplete qos to bootstrap service recommendation, *Proceedings of the 19th IEEE International Conference on Web Services (ICWS 2012)*, Honolulu, HI, USA, June 24-29, 2012, 194-201.

Yu, Q., Zheng, Z., Wang, H. (2013). Trace norm regularized matrix factorization for service recommendation, *Proceedings of the 20th IEEE International Conference on Web Services (ICWS 2013)*, Santa Clara, CA, USA, June 28 - July 3, 2013, 34-41.

Su, X., Khoshgoftaar, T. M. (2009). A survey of collaborative filtering techniques, *Advances in Artificial Intelligence*, 2009, Article ID 421425.

Cohen, W. W., Schapire, R. E., Singer, Y. (1997). Learning to order things, *Proceedings of the 11th Conference on Advances in Neural Information Processing Systems (NIPS 1997)*, Denver, CO, USA, December 1-6, 1997, 451-457.

Liu, N. N., Yang, Q. (2008). Eigenrank: a ranking-oriented approach to collaborative filtering, *Proceedings of the 31st Annual International ACM SIGIR Conference on Research and Development in Information Retrieval (SIGIR 2008)*, Singapore, July 20-24, 2008, 83-90.

Yang, C., Wei, B., Wu, J., Zhang, Y., Zhang, L. (2009). CARES: a ranking-oriented CADAL recommender system, *Proceedings of the 9th ACM/IEEE-CS Joint Conference on Digital Libraries (JCDL 2009)*, Austin, TX, USA, June 15-19, 2009, 203-212.

Liu, Q., Chen, E., Xiong, H., Ding, C.H.Q., Chen, J. (2011). Enhancing Collaborative Filtering by User Interest Expansion via Personalized Ranking, *IEEE Transactions on Systems, Man, and Cybernetics, Part B: Cybernetics*, 42(1), 218-233.

Balakrishnan, S., Chopra, S. (2012). Collaborative ranking, *Proceedings of the 5th International Conference on Web Search and Web Data Mining (WSDM 2012)*, Seattle, WA, USA, February 8-12, 2012, 143-152.

Adomavicius, G., Kwon, Y. (2012). Improving Aggregate Recommendation Diversity Using Ranking-Based Techniques, *IEEE Transactions on Knowledge and Data Engineering*, 24(5), 896-911.

Zheng, Z., Lyu, M. R. (2013). Ranking-Based QoS Prediction of Web Services, *QoS Management of Web Services*, Zhejiang University Press, Hangzhou, China, 83-96.

Bell, R. M., Koren, Y. (2007). Scalable collaborative filtering with jointly derived neighborhood interpolation weights, *Proceedings of the 7th IEEE International Conference on Data Mining (ICDM 2007)*, Omaha, Nebraska, USA, October 28-31, 2007, 43-52.

Takács, G., Pilászy, I., Németh, B., Tikk, D. (2007). Major components of the gravity recommendation system, *ACM SIGKDD Explorations Newsletter*, 9(2), 80-83.

Marden, J. I. (1996). *Analyzing and modeling rank data*, CRC Press, Boca Raton, Florida, USA.

Salakhutdinov, R., Mnih, A., Hinton, G. (2007). Restricted Boltzmann machines for collaborative filtering, *Proceedings of the 24th International Conference on Machine learning (ICML 2007)*, Corvallis, Oregon, USA, June 20-24, 2007, 791-798.

Blei, D. M., Ng, A. Y., Jordan, M. I. (2003). Latent dirichlet allocation, *The Journal of machine Learning Research*, 3, 993-1022.

Gorrell, G., Webb, B. (2005). Generalized Hebbian algorithm for incremental latent semantic analysis, *Proceedings of the 9th European Conference on Speech Communication and Technology (EUROSPEECH 2005)*, Lisbon, Portugal, September 4-8, 2005, 1325-1328.

Netflix Update: Try This at Home. (2006). Retrieved January 12, 2015, from http://sifter.org/~simon/journal/20061211.html.

Zheng, Z., Zhang, Y., Lyu, M. R. (2010). Distributed QoS Evaluation for Real-World Web Services, *Proceedings of the 17$^{th}$ International Conference on Web Services (ICWS 2010)*, Miami, Florida, USA, July 5-10, 2010, 83-90.

Zhang, Y., Zheng, Z., Lyu, M. R. (2011). Exploring Latent Features for Memory-Based QoS Prediction in Cloud Computing, *Proceedings of the 30$^{th}$ IEEE Symposium on Reliable Distributed Systems (SRDS 2011)*, Madrid, Spain, October 4-7, 2011, 1-10.

Järvelin, K., Kekäläinen, J. (2002). Cumulated gain-based evaluation of IR techniques, *ACM Transactions on Information Systems*, 20(4), 422-446.

Resnick, P., Iacovou, N., Suchak, M., Bergstrom, P., Riedl, J. (1994). Grouplens: An Open Architecture for Collaborative Filtering of Netnews, *Proceedings of the 5$^{th}$ ACM Conference On Computer Supported Cooperative Work (CSCW 1994)*, Chapel Hill, NC, USA, October 22-26, 1994, 175-186.

Paterek, A. (2007). Improving regularized singular value decomposition for collaborative filtering, *Proceedings of KDD Cup and Workshop 2007*, San Jose, California, USA, August 12, 2007, 39-42.

Kiwiel, K. C. (2001). Convergence and efficiency of subgradient methods for quasiconvex minimization, *Mathematical Programming (Series A)*, 90(1). 1-25.

Kiwiel, K. C. (2003). Convergence of approximate and incremental subgradient methods for convex optimization, *SIAM Journal of Optimization*, 14(3), 807–840.

Robbins, H., Siegmund, D. (1971). A convergence theorem for non negative almost supermartingales and some applications, *Proceedings of a Symposium Held at the Center for Tomorrow*, the Ohio State University, Columbus, Ohio, USA, June 14-16, 1971, 233-257.